# Justifying the Dependability and Security of Business-Critical Blockchain-based Applications

Pierre-Yves Piriou
*Dpt. PRISME*
*EDF Lab*
Chatou, France

Olivier Boudeville
*Dpt. PERICLES*
*EDF Lab*
Saclay, France

Gilles Deleuze
*Dpt. PERICLES*
*EDF Lab*
Saclay, France

Sara Tucci-Piergiovanni
Université Paris Saclay,
CEA, LIST
F-91120, Palaiseau, France

Önder Gürcan
Université Paris Saclay,
CEA, LIST
F-91120, Palaiseau, France

*Abstract*—In the industry, blockchains are increasingly used as the backbone of product and process traceability. Blockchain-based traceability participates in the demonstration of product and/or process compliance with existing safety standards or quality criteria. In this perspective, services and applications built on top of blockchains are business-critical applications, because an intended failure or corruption of the system can lead to an important reputation loss regarding the products or the processes involved. The development of a blockchain-based business-critical application must be then conducted carefully, requiring a thorough justification of its dependability and security. To this end, this paper encourages an engineering perspective rooted in well-understood tools and concepts borrowed from the engineering of safety-critical systems. Concretely, we use a justification framework, called CAE (Claim, Argument, Evidence), by following an approach based on assurance cases, in order to provide convincing arguments that a business-critical blockchain-based application is dependable and secure. The application of this approach is sketched with a case study based on the blockchain HYPERLEDGER FABRIC.

## I. INTRODUCTION

As a concept, blockchains are expected to uphold applications with valuable properties such as data immutability, transparency and accountability in many areas such as finance, industry and public services. As the properties of blockchains lead to use cases that involve critical business missions such as certification, financial exchanges or traceability, many blockchain-based systems are considered "business-critical systems", whose Dependability and Security (D&S) must be thoroughly justified from the design to the operational phases to decision-makers including senior management and regulators. Usually, dependability engineering is used to deal with *safety*-critical systems (like transportation, aerospace, ...) whose failure can have a significant impact on people, environment or assets, all of which are valuable in the physical world. By analogy, a *business*-critical system performs missions whose failure can have a significant impact on the intangible assets (like financial, legal, reputation ...) of an organisation; for this reason dependability engineering is also relevant for these systems.

Unfortunately, there is currently a lack of standards that can be used as a guide for engineering business-critical systems based on blockchains. Numerous published works contribute to strengthen the D&S of specific blockchains protocols. On the one hand, formal works are conducted to prove desirable properties on the distributed protocols (like the Byzantine fault tolerance of consensus algorithms [4] or the safety and liveness of transaction validation [7]). On the other hand, analyses are performed to identify specific risks and investigate mitigation solutions (e.g., [12] for Bitcoin, [8] for Hyperledger Fabric, [25] for public blockchains...). Nevertheless it seems not trivial to determine how this collection of evidences can be put together to build an end-to-end justification that a business-critical application relying on a blockchain system meets its high-level D&S requirements.

Therefore, this paper addresses the following issue: *How a collection of risk mitigation measures can be organized into an argumentation justifying the D&S of a blockchain application?* and proposes a dedicated approach as a contribution.

The proposed approach is inspired by "assurance (security and safety) cases" [6], [14], coming from nuclear and aerospace industries [5], [9], [19], which is a structured argument, supported by evidences, intended to justify that a system is acceptably assured relative to a concern (such as reliability or security) in the intended operating environment. In a nutshell, the argumentation proposed in this paper consists in decomposing the system under study into functional elements representing the elementary services it should deliver. D&S justification is then conducted via the provisioning of arguments and evidences documenting how the identified risks have been mitigated. To structure this argumentation, we adopt the Claim-Argument-Evidence (CAE) framework [6], [11].

We believe that this paper will help paving the way for industrial engineering of blockchain-based systems, thus promoting their use and acceptability.

The paper is organised as follows: section II develops the industrial motivation and the related works. Section III proposes a general guideline to engineer a dependable and secure blockchain based system, introducing in particular a CAE template for blockchain based applications. Section IV instantiates the CAE template considering a fictive application based on the HYPERLEDGER FABRIC blockchain. Finally, section V concludes this paper by introducing our ongoing works on dependable engineering and uses of blockchain based applications.

## II. INDUSTRIAL MOTIVATION AND RELATED WORKS

Regulated industrial activities (such as nuclear) imply a reliable and secure traceablility of quality convincing data collected during the life-cycle of safety-critical equipment (fabrication, qualification, maintenance...). These data are distributed into all actors of the industrial sector (integrators, suppliers, subcontractors...) what makes difficult to ensure their long time integrity and consistency. Trust on these data are generally provided by costly third-party certification procedures ensuring only a partial arbitrary coverage. Blockchain technologies (or more generally distributed ledgers) promise the integrity and availability of the registered data, while allowing a control of the data through smart-contract ensuring its compliance and consistency. These properties are a priori relevant arguments to gain trust of regulators on a certification-aided system based on Blockchain technologies [23]. Since the targeted data should prove the quality of safety-critical equipments, the registering system is business-critical. Therefore to be ultimately acceptable by regulators, the D&S of such system should be thoroughly justified (*Dependability is the ability to deliver a service that can **justifiably** be trusted* [3]).

In the literature, several studies contribute to strengthen the D&S of a blockchain-based application, by analysing the risks of specific blockchain implementation and proposing mitigation measures [1], [8], [12], [15], [20], [24], [25]. These works provide a large number of factual pieces of evidence from different natures (simulation, demonstration, testing, statistical analysis, formal analysis etc.) contributing to build the assurance case. Nevertheless to manage the complexity of the whole system, we claim that there is a need for a framework to build upon them the argumentation capturing the whole justification chain of the top-level claim: "the application is dependable and secure" without any missing links. The notion of justification framework refers to structuring and capitalizing the reasoning chain that has been followed throughout the design process of a complex system to provision the D&S (by ensuring that there is a clear link between D&S claims and D&S designations). Such approach is called "Assurance (security and safety) cases" in [6], which are defined as "documented bodies of evidence that provide valid and convincing arguments that a system is adequately dependable in a given application and environment" [19].

The two main identified justification frameworks are the Goal Structuring Notation (GSN) [16], and the Claim-Argument-Evidence (CAE) [5], [11] which are highly recommended by safety-critical systems regulators [19] and is standardized in [14]. The main difference between both frameworks resides in the characterization of arguments: GSN uses a generic notion of *argument strategy*, whereas CAE introduces three argument types: the *decomposition*, the *substitution* and the *concretization*. Hence CAE can be seen as a refinement of GSN. For this reason, we adopt the CAE framework in this paper.

## III. ON THE ENGINEERING OF DEPENDABLE AND SECURE BLOCKCHAIN BASED SYSTEMS

This section proposes a guideline to engineer a business-critical blockchain-based system while justifying the fulfillment of its D&S requirements. It mainly refines the usual system engineering workflow of mission-critical system design for blockchain, which entails functional analysis: breaking-down the system in smaller functional elements.

### A. Functional analysis

D&S requirements refer to the expected functionalities of the system. First step is then to address its functional analysis. Figure 1 illustrates the fundamental functions of a blockchain system: as a distributed ledger system it must asynchronously handle *read* and *write* operation requests. To be eventually registered following a *write* operation, data must be encapsulated in transactions that must be syntactically correct and semantically *valid*. Validity criteria depend both on the system specification (e.g. a transaction must be signed by a valid signature) and on the application needs which are usually enforced by smart contracts (e.g. a numerical value extracted from the transaction data must be in a specific range). The yellow note on the Figure 1 gives an example of validity criteria matching the token-based blockchain protocol specifications. Moreover, as a distributed system, a blockchain is expected to guarantee consistency: users should read consistent values despite the underlying distribution (at this stage of the system description, consistency can be described in an intuitive way and refined later through formal specifications [22]). To summarize, as a distributed ledger system, blockchain *functional elements* (FE) are specified as follows:

- FE1: register any valid transaction eventually
- FE2: register only valid transactions
- FE3: answer consistently to *read* requests

These fundamental FEs can be easily mapped to classical requirements of distributed systems: FE1 satisfies a *liveness* requirement while FE2 and FE3 satisfy *safety* requirements (*validity* and *consistency*). It is important to notice that some blockchains may weaken their service with respect to these FEs (and associated requirements). Typically, public blockchains may occasionally fail to register valid transactions or temporarily fail to guarantee consistency under degraded network conditions. Given a blockchain protocol and an application logic, the validity criteria of transactions and the consistency criteria of read results should be defined. Moreover, depending on the application, additional functional elements may be expected. For example, if the system should protect its confidentiality then specific functional elements should be put in place.

### B. Risk preliminary analysis

The second step of the engineering process is to list the risk of faults and failures and classify them against their criticality and their likelihood. According to the functional analysis, feared events that we want to avoid here could be:

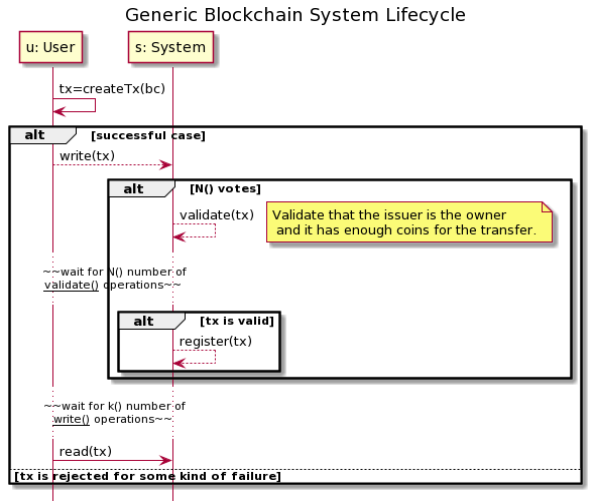

Fig. 1. Generic functional analysis of a blockchain-based application

1) registering of an invalid transaction
2) rejection or deletion of a valid transaction
3) inconsistent read result

Such failures can be refined to match better the application services. For example, the main feared event for financial applications is the famous *double spending*, which is an event that compromises consistency. Indeed, if the system is not well-designed or flawed, an attacker can register two conflicting transactions.

Such failures are induced by the propagation of errors in the system that are activated in unfavorable conditions because of the presence of elementary faults [3]. Such risks of faults should also be identified to the best of our knowledge. As a computing system, a blockchain is subject to all classical risks of faults affecting both software and hardware parts (due to poor quality of development, malicious activities, physical hazard...). But because blockchains are designed to be trustworthy, they address by nature workflows where some actors would have an interest in behaving incorrectly. It is therefore often very relevant to focus on Byzantine faults when analysing the risks of a blockchain application [18]. In particular, *fraud* and *censorship* are two kinds of malicious faults that can be injected by an attacker attempting to provoke respectively the above mentioned feared events 1 and 2.

Additional non-malicious faults should also be thoroughly considered, depending on the protocol specification and the application logic. Typically, some blockchain applications suffer from performance issues, so that when the system is overloaded (too many transactions to handle in parallel), some valid transactions are unreasonably delayed or even aborted. This kind of failure results from non-malicious faults (due to a poor quality of development of the system or its misuse during the operational stage): inadequacy of the system performance to scale-up for the application needs.

*C. Risk mitigation*

More critical risks of fault should be mitigated by appropriate measures taken during engineering phase or operational phase. According to [3], there are four categories of risk mitigation measures:

- **fault prevention**: the application of best practice during the development stage is a simple way that is often enough to prevent many basic faults. Such recommendations are often supplied in the (official or side) documentation of a computing system (including blockchain based). Fault prevention measures can also be deployed during the operational stage, e.g. by limiting the system operation to permissioned users (for consortium blockchains like HYPERLEDGER FABRIC [2]).
- **fault elimination**: this is the main purpose of the verification stage in the development process. Development faults can be detected then removed using dynamic methods where the system is experienced (testing, symbolic execution). Furthermore, to prove that an implementation is free of fault, formal static methods may be exploited (static analysis, deductive verification, model-checking). During the operational phase, maintenance measures can be deployed to eliminate current faults.
- **fault tolerance**: this kind of mitigation measures is generally provided by design. Fault tolerance is obviously a natural key feature of distributed system. They can be as simple as replication on redundant architecture or based on more complex techniques to detect, diagnose and handle errors and faults in order to recover an erroneous system. But to take advantage of a fault tolerant design, a particular attention should be applied in the system configuration. For example, if a byzantine fault-tolerant consensus algorithm is used in a too little architecture of only 3 nodes, the system has a single point of failure, despite the fact that it is theoretically fault-tolerant. Indeed, such algorithm can tolerate, by nature, a proportion of byzantine nodes strictly inferior than 1/3 [18]. The configuration of fault-tolerant complex system can be guided by simulation techniques.
- **fault forecasting**: once a set of measures have been taken to prevent, eliminate and tolerate a large diversity of faults, the residual risks can be analysed by qualitative or quantitative assessment to check that the D&S requirements has been met. Simulation technique can once again be exploited to perform such kind of analysis.

*D. Justifying the D&S using a CAE tree*

Because the D&S of a critical system must by definition be justified to be accepted [3], this section introduces a framework to structure a justification argumentation. A justification of a claim is a structured collection of arguments built upon elementary evidences. This structure can be represented as a tree whose root is the claim to justify, the intermediate nodes are steps of argumentation and the leaves are elementary facts. Such formalism is called a Claim-Argument-Evidence (CAE) tree [5].

Three kinds of argumentations can be used to refine a high level claim into several focused and less ambiguous claims:

- **Decomposition**: a claim is decomposed in a conjunction of simpler subclaims. When it is not trivial, the validity of the inference relation gives rise to an auxiliary subclaim. For example, a claim on a computing system may be decomposed into its software and hardware components.
- **Substitution**: a claim related to a given object is transposed into the analogous claim related to an equivalent object. Once again, when it is not trivial, the validity of the equivalence relation gives rise to an auxiliary subclaim. For example a claim on the target system may be substituted for a claim on an equivalent simulated system.
- **Concretization**: an abstract claim is refined by introducing for example a definition or a quantified value.

Finally, evidences may be for instance facts like design choices preventing or tolerating by nature some faults or results of analysis performed to eliminate or forecast some other faults. There are two categories of evidences:

- **Hypothesis**: what should be commonly accepted
- **Proof**: what cannot be opposed

Figure 2 shows a template of CAE tree that can be instantiated to justify the D&S of a given blockchain based application (in our graphical representation, claims, arguments and evidences are respectively colored in blue, yellow and green. Evidences are introduced by their type "Hypothesis" or "Proof", colored in orange). The first argument is a decomposition justified by the general functional analysis made in subsection III-A. For a particular application built upon a specific blockchain protocol, the corresponding validity criteria of transactions and the consistency criteria of read queries have to be defined. Finally, any risks identified during the risk analysis affecting one of the Functional Elements (FE) must be addressed either by providing evidence of their mitigation measures, or by justifying their acceptability. Note that this level of abstraction is independent of the type of blockchain.

## IV. CASE STUDY: JUSTIFYING THE D&S OF A HYPERLEDGER FABRIC BASED APPLICATION

This section instantiates the CAE template represented on Figure 2 for an application built upon the HYPERLEDGER FABRIC blockchain [2]. The application is fictive and will not be described since the aim here is not to discuss the relevance of a dependable design for a specific application but to demonstrate the application of the CAE method to a given blockchain.

### A. Refining the functional analysis

Subsection III-A introduces three generic functional elements (FE1, FE2 and FE3) of a blockchain-based application that should be refined by specifying the validity criteria of the transactions and the consistency criteria of the read results. This subsection applies this refinement for an HYPERLEDGER FABRIC based application.

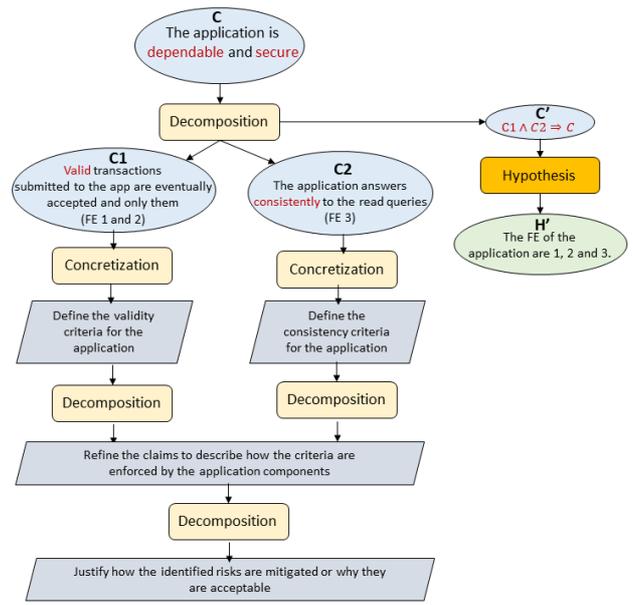

Fig. 2. CAE template to justify the D&S of a blockchain-based business-critical application.

*1) Validity criteria:* Seven criteria of validity are checked in turn whenever a transaction is to be processed. They can be split into two sets, each one being enforced by a different role in terms of participants to the protocol.

First, in the course of the execution phase, the endorsers have to assess whether three properties are met:

- **V1 - the transaction emitter is legit**: its signature has been recorded by the Membership Service Provider (MSP) and it belongs to the emitter set.
- **V2 - the business logic is respected**: this point is validated by a successful execution of the application chaincode
- **V3 - the transaction is unique**: i.e. it has not already been acknowledged (as checked with a nonce-based anti-replay mechanism)

If these three conditions are met, the endorser peers endorse the transaction by sending its effects on the database (read and write sets) to the application's client. The reply is signed by the endorser. Once the client has received the endorsements, it can submit the transaction to the orderer service. The orderers execute a consensus algorithm to commit a new block, deciding on an order for the ongoing transactions. During this phase, no validity criterion is checked.

Finally, once any peer receives a new block, it performs the validation phase, checking whether:

- **V4 - the endorsement policy has been respected**: it complies with the governance rules (which takes part of the application configuration)
- **V5 - the endorsers are legit**: their signatures have been recorded by the Membership Service Provider (MSP) and they belong to the endorsers set.

- **V6 - the answers of the endorsers are consistent**: they have computed the same effects on the database
- **V7 - the transaction is not in conflict with another one already applied**: two transactions are in conflict if their effects on the database are conflictual (e.g. if the two transactions would write on the same key). This checking follows the Multiversion Concurrency Control (MVCC) method commonly used by database management systems.

If these 4 conditions are met, the peers tag the transaction *valid* and apply its effects on their local copy of the database (this means that the transaction has been accepted by the system).

The CAE tree extract drawn on the Figure 3 shows how the claim **C1** (from Figure 2) can be concretized then decomposed into subclaims related to the related HYPERLEDGER FABRIC components. First, the abstract notion of *validity* is concretized through the seven criteria listed above, resulting in the claim **C1c**. This claim is then decomposed into 5 subclaims. The subclaims **C1c.1** and **C1c.4** are straightforward since they introduce the two roles that check the validity criteria. The subclaim **C1c.2** introduces the MSP component since it is used by the endorsers and the peers to check respectively the criteria V1 and V5. The subclaim **C1c.3** introduces the orderer components because, although they do not apply any validation criteria, the committing of blocks is a prerequisite to the checking of criteria V4 to V7 by the peers. Finally, the subclaim **C1c.3** introduces the communication network because all these components communicate by message passing. The hypothesis **H1c'** justifies this decomposition by defining what means exactly the acceptation of a transaction.

*2) Consistency criteria:* Every peers in the system maintain a local replica of the application database (a key-value store). To answer consistently to read requests on this database, they have to synchronize their local states. This is done through the blocks of transactions committed by the orderer service and propagated to the network using a gossip protocol. Then as we can see on Figure 4, the claim **C2** can be concretized by introducing the criterion of consistency **C2c**: Two correct peers synchronized on the same block have the same state of their local key-value store. To achieve this criterion HYPERLEDGER FABRIC uses three kind of components, following an *Execute-Order-Validate* paradigm:

- **Execute:** application's transactions are firstly executed in parallel by *endorser peers* considering the current blockchain state. This execution may fail for different reasons (detailed in the subsection IV-A1) and computes the effects of the transaction on the application database state but does not apply them.
- **Order**: pending transactions are ordered in a new block by *orderer peers* executing a raw consensus algorithm (no validity criterion of transactions is checked during this phase)
- **Validate**: all *peers* apply the *valid-after-ordering* transactions to update their local copy of the database (cf. subsection IV-A1).

Thus, assuming that the local databases of all correct peers are initialized with the same initial state (**H2c'**), if we can justify that the three involved sets of peers (endorsers, orderers and standard peers) are able to deliver their corresponding services (**C2c.1**, **C2c.2** and **C2c.3**), then we justify the claim **C2c** and therefore the claim **C2**.

*B. Risk analysis and mitigation instantiation examples*

The refining of functional analysis results in several subclaims concerning the elementary services of the blockchain's components. To justify these subclaims, we have to demonstrate that any identified risk of faults affecting these components have been mitigated and that the residual risks are acceptable. What follows exemplifies this process for the subclaims **C1c.1** and **C2c.2**. Several published works identify generic risks regarding the HYPERLEDGER FABRIC blockchain ( [1], [8]), but they can be completed or refined to fit better with a specific application (for instance, some applications do not have particular concerns about privacy). Note that the purpose here is not to discuss the relevance of the risk analysis and the mitigation measures but only to illustrate how can they contribute to the justification of higher-level claims.

*1) Risks affecting the endorsers:* Assuming that the risk analysis for the endorsement part of the transactions validation identifies the following list of possible faults:

1) Bugs in the implementation of the endorsement functionality of the peers: design fault in HYPERLEDGER FABRIC code.
2) Bugs in the implementation of the application's chaincode executed by the endorsers: design fault in the business logic implementation.
3) Crashes of endorser peers: hazard during operation
4) DoS attack on endorser peers: some internal participant may have interest to impede some valid transactions.
5) Fraud of endorser peers: some organizations involved in the endorsement of transaction may have interest to behave wrongly by fraudulently accepting an invalid transaction.
6) Censorship of endorser peers: some organizations involved in the endorsement of transaction may have interest to behave wrongly by censoring (ie rejecting) a valid transaction.

The CAE represented on Figure 5 shows how the mitigation of these risks of faults can be justified. The first kind of faults is eliminated by testing methods applied by the Hyperledger foundation. The System Verification Test (SVT) report is used here as a proof[1] (**P1c.1.1**). The second kind of faults is prevented by the application of formal methods to guarantee the absence of flaw and the correctness of the implementation with regard to a formal specification (**P1c.1.2**). Finally, the four other faults are mitigated by a fault-tolerant endorsement policy, which determines the quality and quantity of endorsements that a client of the application has to collect in order to validate its transaction proposal (cf. the criterion of validity V4). The endorsement policy should be wisely configured to find a compromise between the tolerance of the different risks identified. Indeed, extreme policies cannot tolerate all kinds of faults. A policy requiring that all endorsers sign the transactions tolerates very well the attempts of frauds but a single malicious endorser can easily censor a transaction. On the contrary, a policy requiring transactions to be signed by any endorser tolerates very well the attempts of censorship but a single malicious endorser may bypass the chaincode execution to fraudulently endorse an invalid transaction. For a particular application, depending on actual participants and trust relations between them, a compromise should be found, possibly resulting in a custom endorsement policy. The determination of such a fault tolerant policy can be guided by a simulation-based approach. The report of this simulation analysis is used in the justification to establish that the residual risks are acceptable (**P1c.1.3**).

---
[1]This report and the testing methodology can be consulted at https://wiki.hyperledger.org/display/fabric/Quality+Assurance\%3A+Tests\%2C+Strategy\%2C+Reports (visited on 15/06/21)

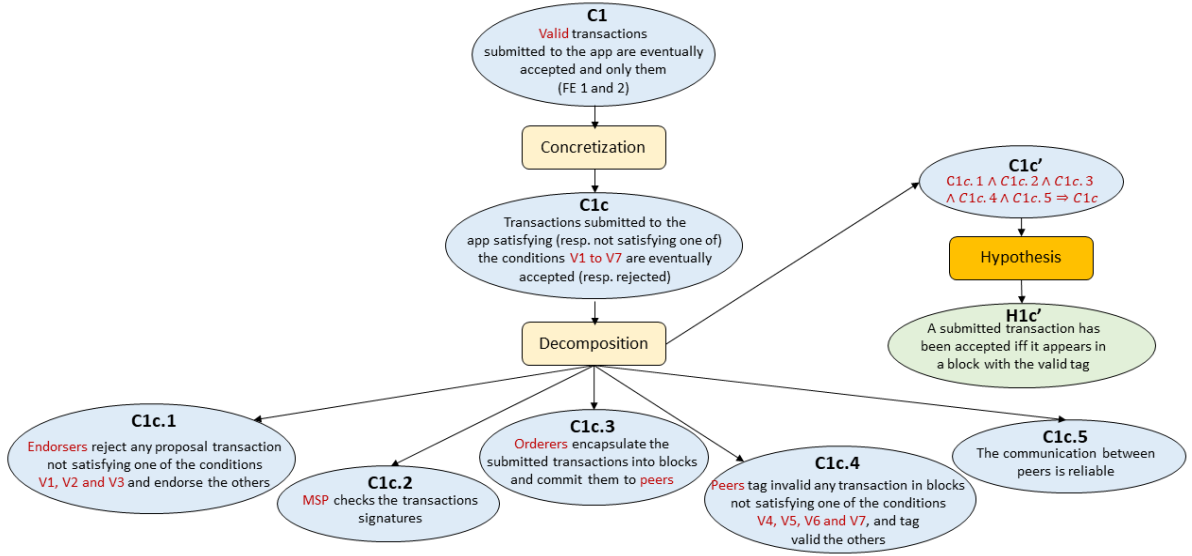

Fig. 3. CAE concretizing the validity criteria for HYPERLEDGER FABRIC

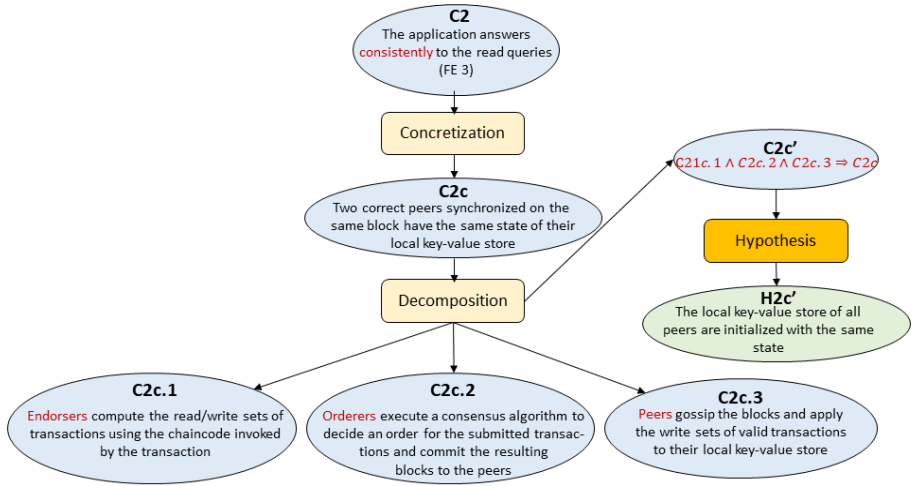

Fig. 4. CAE concretizing the consistency criterion for HYPERLEDGER FABRIC

*2) Risks impacting the orderers:* Let us consider now that the designer of the target application has chosen to use the consensus algorithm Raft (HYPERLEDGER FABRIC supports several consensus algorithms). This algorithm is Crash-Fault Tolerant (CFT) [21]. This implies in particular that the fault analysis would not identify any risks of Byzantine fault of orderer peers (for example, we could argue that in a permissioned blockchain, orderers have no rational interest in attacking the consensus). However figure 6 shows a CAE justifying the subclaim **C2c.2**. First the claim is substituted by the claim **C2c.2s** replacing the generic consensus algorithm by the particular Raft algorithm. Then two mitigation measures are used to justify that Raft allows to deliver the ordering service, assuming the risk analysis conclusions (**H2c.2s'**):

- Bugs in etcd (the Raft implementation embedded in HYPERLEDGER FABRIC) have been eliminated (**C2c.2s.1**) by functional tests performed by its developer[2] (**P2c.2s.1**).
- Design faults of Raft have been prevented (**C2c.2s.2**) by a formal proof provided by its author [21] (**P2c.2s.2**).

*C. Discussion*

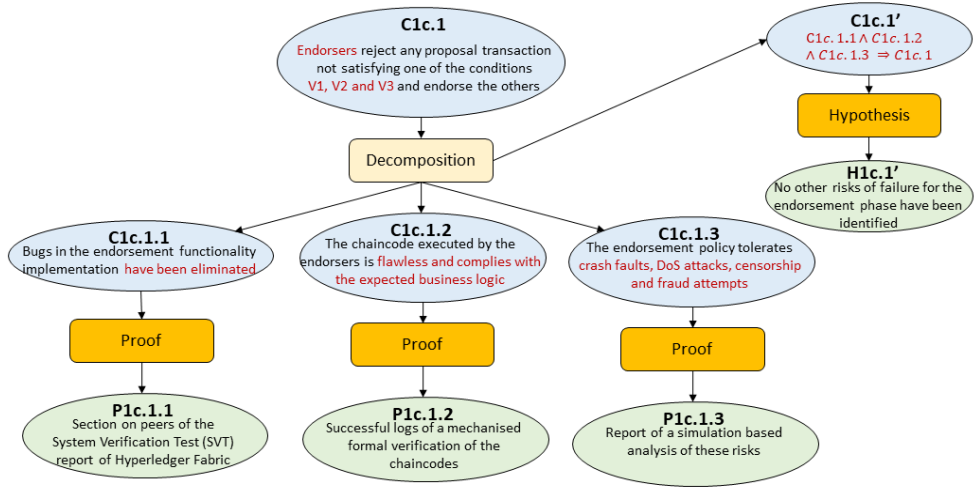

Fig. 5. CAE justifying the subclaim C1c.1

We found that a CAE tree is a relevant framework to build confidence regarding the D&S of a blockchain-based application. Indeed:

- it enforces a well-structuring of the argument, which is a required condition for managing the complexity of such systems,
- it supports discussion and reduces time-to-agreement on what evidence is needed and what the evidence means,
- having established the argument structure soon in the engineering process, it focuses D&S activities towards the lacking evidences,
- it enables the recognition of convincing argument patterns and then supports monitoring of project progress towards successful qualification in the prospect of a regulator acceptance.

Our case study is based on a permissioned blockchain because they are natural candidates to support industrial business-critical systems. Nevertheless, we believe that our approach is generic enough to be applicable for permissionless blockchains, where risk mitigation measures should focus more on the operational phase than on the engineering phase of the considered application. The main difference would therefore be on the nature of evidences that would be provided more by probabilistic assessment than by deterministic analysis. These intuitions have to be verified in future works.

## V. CONCLUSION AND PROSPECTS

This paper takes advantage of a justification framework, called Claim-Argument-Evidence [5], to build a convincing argument that a business-critical blockchain-based systems is dependable and secure. This approach is inspired by assurance cases that traditionally involve safety-critical systems. The framework is applied on a fictive use case based on the HYPERLEDGER FABRIC blockchain. Such an approach requires a preliminary D&S engineering of the targeted application, where risks are identified and mitigation measures are applied, to prevent, eliminate, tolerate or forecast them. Since this approach is encouraged for demonstrating the D&S of a complex critical system to a regulator auditor, we hope that our work will favor blockchain acceptability for regulated industrial applications.

For future work, we plan to apply our approach on real use cases (addressing the certification of industrial operations such as tensile testing or welding [23]) and to develop specific mitigation measures for justifying its D&S. We are contributing in particular to the formal verification of HYPERLEDGER FABRIC smart contracts, and the multi-agent simulation of blockchain systems [13] (using respectively the tools Why3 [10] and MAX [17]).

---

[2]cf. https://github.com/etcd-io/etcd/tree/main/tests/functional (visited on 15/06/21)

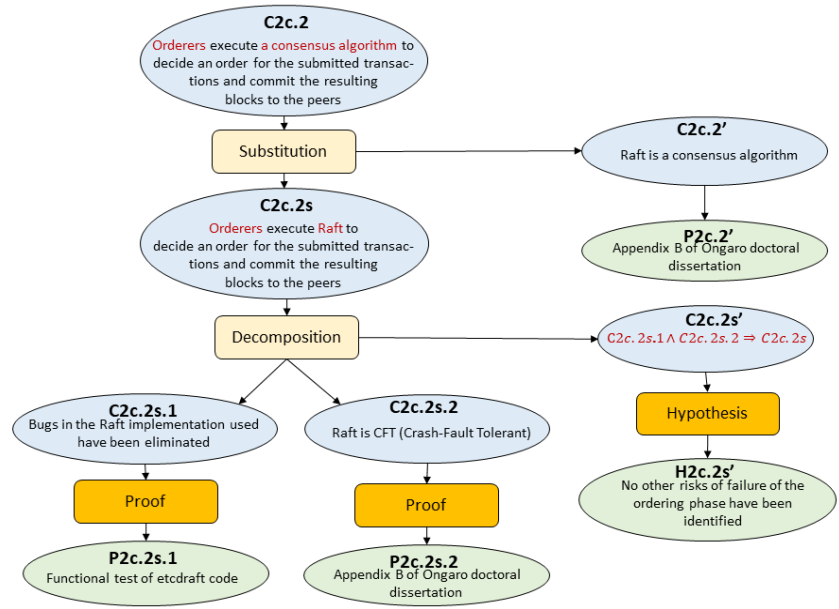

Fig. 6. CAE justifying the subclaim C2c.2